\documentclass[%
 reprint,
 amsmath,amssymb,
 aps,
]{revtex4-2}

\usepackage{graphicx}
\usepackage{dcolumn}
\usepackage{bm}

\begin{document}

\preprint{APS/123-QED}

\title{Polychromatic Electric Field Knots}
\author{Manuel F. Ferrer-Garcia$^{1}$}
\author{Alessio D'Errico$^{1}$}%
\author{Alicia Sit$^{1}$}
\author{Hugo Laroque$^{2}$}
\author{Ebrahim Karimi$^{1}$}
 \email{ekarimi@uottawa.ca}
\affiliation{$^{1}$Department of Physics, University of Ottawa, Advanced Research Complex, 25 Templeton, Ottawa ON Canada, K1N 6N5}%

\affiliation{$^{2}$Research Laboratory of Electronics, Department of Electrical Engineering and
Computer Science, Massachusetts Institute of Technology, Cambridge, MA, USA.}%

\date{\today}

\begin{abstract}
The polarization of a monochromatic optical beam lies in a plane, and in general, is described by an ellipse, known as the polarization ellipse. The polarization ellipse in the tight focusing (non-paraxial) regime forms non-trivial three-dimensional topologies, such as M\"obius and ribbon strips, as well as knots. The latter is formed when the dynamics of specific polarization states, e.g., circular polarization states, is studied upon propagation. However, there is an alternative method to generate optical knots: the electric field's tip can be made to locally evolve along a knot trajectory in time. We propose an intuitive technique to generate and engineer the path traced by the electric field vector of polychromatic beams to form different knots. In particular, we show examples of how tightly focused beams with at least three frequency components and different spatial modes can cause the tip of the electric field vector to follow, locally, a knotted trajectory. Furthermore, we characterize the generated knots and explore different knot densities upon free-space propagation in the focal volume. Our study may provide insight for designing current densities when structured polychromatic electromagnetic fields interact with materials.
\end{abstract}
\maketitle
The rapid advancements in the manipulation and control of electromagnetic radiation has allowed researchers to explore solutions of Maxwell's equations possessing rich topological features. The first structures that became the subject of extensive research were phase~\cite{nye1974dislocations,soskin1997topological} and polarization singularities~\cite{dennis2009singular}. In particular, beams carrying a non-zero value of orbital angular momentum (OAM)~\cite{allen1992orbital} became of interest in many applications of classical and quantum optics. Optical beams with a well-engineered spectrum, polarization, spatial and temporal structures are nowadays widely used in optical manipulation~\cite{he1995direct}, microscopy~\cite{chen2013imaging, zhanghao_super-resolution_2019,hell1994breaking}, surface and material structuring~\cite{nivas2015direct}, and classical and quantum communication~\cite{willner2015optical,bouchard_quantum_2018}. Moreover, more complicated structures can be observed by analyzing an optical beam that is not at a fixed two-dimensional plane but within a three-dimensional volume. Indeed, it is predicted that the free-space trajectories of field dislocations can form closed loops~\cite{berry2001knotted,dennis2010isolated} with non-trivial topologies, e.g. links and knots~\cite{sugic2018singular}. Recently, by exploiting the technology of spatial light modulators, beams exhibiting these features have been successfully engineered in the laboratory~\cite{larocque2018reconstructing} and proposed as tools to encode prime numbers~\cite{larocque2020optical}.
Advanced tools in nonlinear optics have improved our ability to structure the frequency spectrum of an electromagnetic field, with applications such as optical manipulation and atom trapping~\cite{feng_quantum_2002,andersen2006quantized,bhattacharya_optomechanical_2008,kozyryev_coherent_2018}, and structured electric currents in semiconductors~\cite{sederberg2020vectorized}. Nevertheless, the polarization state of polychromatic fields has barely been studied by the community. While in the monochromatic case the electric field vector describes an ellipse, more complicated curves occur when multiple waves with different temporal frequencies are superimposed. This new ``zoo'' of polarization states, as mentioned before, remains almost unexplored; the only accurately described cases are the ones in which both fields oscillate in the same plane. If the two frequencies follow the condition $\omega_2=m\,\omega_1$, where $m$ is a positive integer, the local electric field traces a quartic plane curve (Lissajous-like  curves)~\cite{freund_bichromatic_2003}.
\begin{figure}[t]
\centering
{\includegraphics[width=\linewidth]{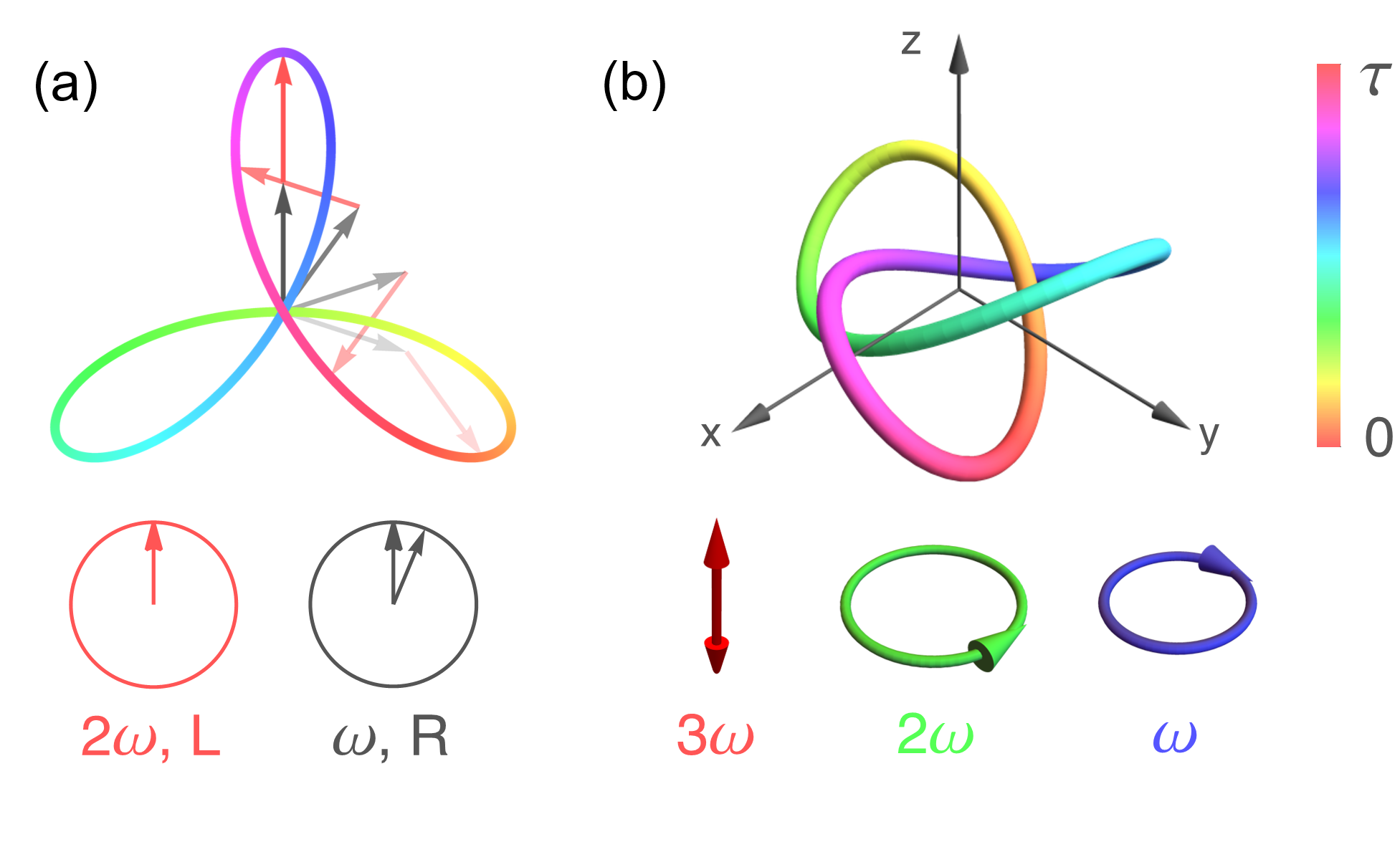}}
\caption{(a) Paraxial polychromatic electric field (tip) tracing the planar quartic polarization curve $\mathcal{C}(t)=[\cos(t)+\sin(2t),-\sin(t)+\cos(2t),0]$. Fundamental ($\omega$) and second harmonic ($2\omega$) beams possess left- and right-handed polarization states, respectively. (b) Non-paraxial polychromatic electric field tracing the three-dimensional parametric curve $\mathcal{C}(t)=[\sin(3t),\cos(t)-2 \cos(2t),\sin(t)+2 \sin(2t)]$. The third harmonic ($3\omega$) is longitudinally polarized.}
\label{fig:fig1}
\end{figure}

In this Letter, we aim to explore the behaviour of structured polychromatic fields in the non-paraxial regime, where a non-negligible component of the electric field along the propagation direction is observed. Non-paraxial fields are typically obtained by focusing paraxial beams with high numerical aperture (NA) lenses~\cite{richards1959electromagnetic}, i.e., in the tight focusing regime. It has been previously shown that, under such conditions, polarization singularities in paraxial fields can be mapped to three-dimensional structures such as M\"obius strips, twisted ribbons and skirmionic textures~\cite{bauer2015observation, bauer2019multi, bauer2020ultrafast,gao2020paraxial}. There has also been hints at how three-dimensional superpositions of plane waves with different frequencies can create an electric field in which its tip locally traces a knotted curve~\cite{sugic2020knotted}. Here, we propose an approach to manipulate and design the polarization structure of tightly focused polychromatic beams in the form of knotted polarization curves by means of vector diffraction of polychromatic beams through an aplanatic lens. Indeed, the knotted polarizations are ubiquitous once simple conditions on the different frequencies are imposed. We follow a heuristic approach to generate large populations of knotted polarizations in the focal plane, based on the knowledge of the tight focusing of vector vortex beams. In the provided examples, we localize and identify the knotted trajectories in different planes in the focal volume.

\begin{figure}[t]
\centering
{\includegraphics[width=\linewidth]{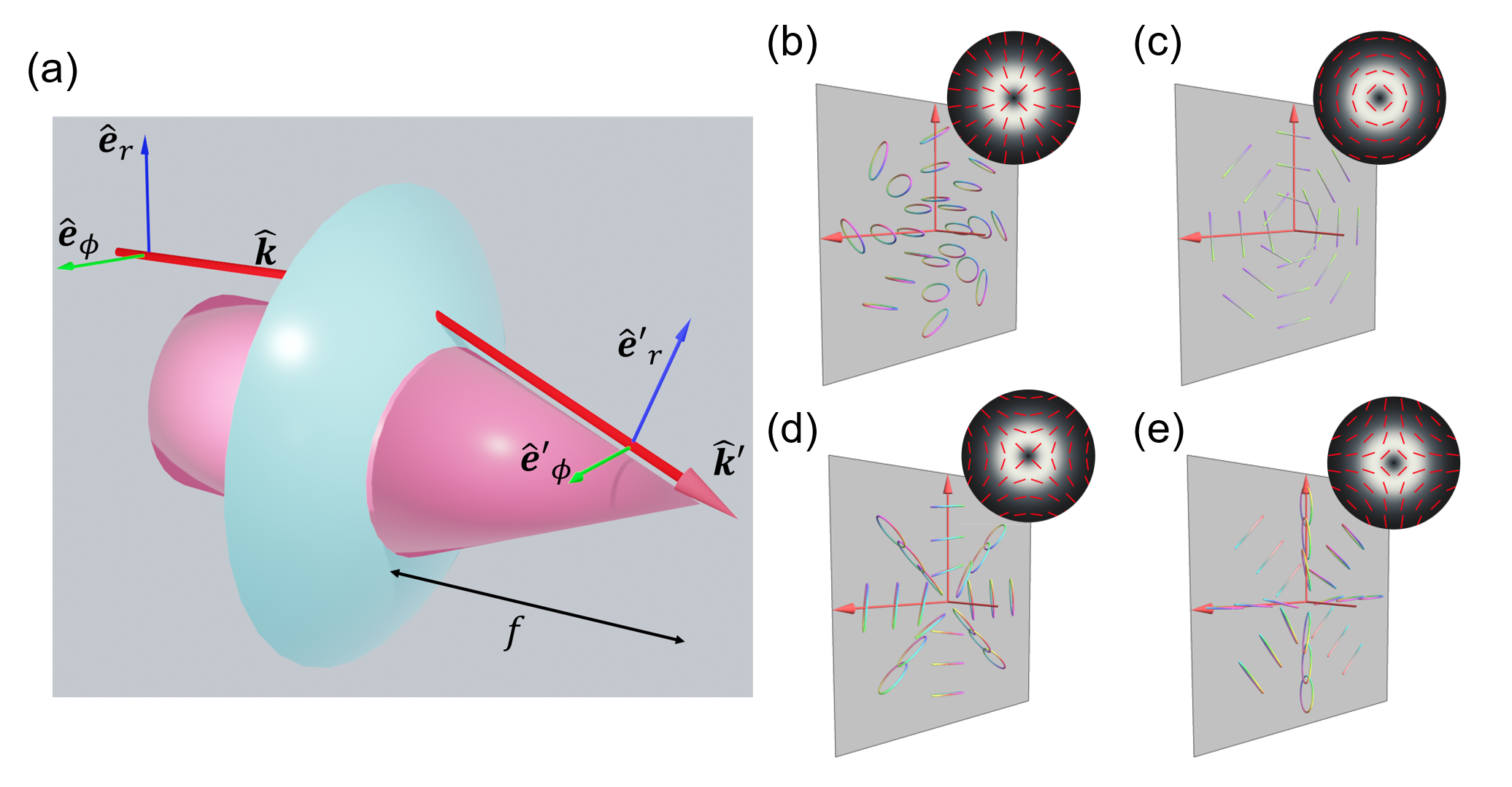}}
\caption{(a) Schematic diagram of the coordinate system transformation used to calculate the tightly focused fields. Three-dimensional polarization distribution at the focal plane $z=0$ for the tightly focused vector vortex beam: (b) $\mathbf{E_v}(\mathbf{r};1,0)$, (c) $\mathbf{E_v}(\mathbf{r};1,\pi)$, (d) $\mathbf{E_v}(\mathbf{r};-1,0)$ and (e) $\mathbf{E_v}(\mathbf{r};-1,\pi)$. Insets show the intensity and polarization distributions of the beams in the paraxial regime.}
\label{fig:fig2}
\end{figure}
%
Let us start by considering a polychromatic electric field described by the superposition of $N$ temporal Fourier components,
\begin{equation}
    \mathbf{E}(\mathbf{r};t)=\sum_{n=1}^N \text{Re}\left[\mathbf{A}_n(\mathbf{r}) e^{-i \omega_n t}\right],
\end{equation}
where $\{\omega_n\}$ is the set of optical frequencies, while $\mathbf{A}_n$ is the local complex amplitude. For each point in space, the tip of the electric field vector describes a closed three-dimensional curve $\mathcal{C}(t)$ for $t\in [0, 2\pi/\omega_p]$, where $\omega_p$ is the greatest common divisor of the set $\omega_{n}$ (see Figure \ref{fig:fig1}). In general, it is possible to construct a set of parametric equations such that $\mathcal{C}(t)= [ X(t),Y(t), Z(t)]$, where $X(t)$, $Y(t)$ and $Z(t)$ are finite Fourier series with $i, j$ and $k$ frequency components.
The parametrized curve $\mathcal{C}(t)$ is defined as a Fourier-$(i,j,k)$ knot \cite{stasiak1998ideal}. These three-dimensional curves simplify to the well-known polarization ellipse for the case of monochromatic light ($N=1$) and to the \textit{Lissajous-like singularities} for the paraxial case, i.e., $Z(t)=0$~\cite{freund_bichromatic_2003}. Similarly, a non-trivial knot is defined as a curve that cannot be transformed through smooth deformations into a simple loop or \textit{unknot}. Knots can thus be  classified in terms of topological invariants, i.e., mathematical objects that label sets of curves which can be smoothly transformed into each other~\cite{kauffman2001knots}. Any transition from one type of knot to another corresponds to passing through a self-intersecting curve. For the simplest case of a Fourier-(1,1,1) knot, the local electric field must have the form,
\begin{equation}
    \mathbf{e}= \sum_{m=1}^{3} \text{Re}\left[ a_m \, e^{-i \omega_m t} \, \mathbf{\hat{e}_m}\right],
\end{equation}
where $\{\mathbf{\hat{e}_m}\}$ is the set of unit vectors that define a local orthogonal three-dimensional basis, and $\{a_m\}$ are complex numbers related to the relative intensity and phase among the chromatic components. In addition, the frequencies $\{\omega_m\}$ must be proportional to three co-prime numbers~\cite{kauffman2001knots}.

For paraxial beams, one can only create two-dimensional polarization curves (e.g., Lissajous figures); this is due to the electric field being confined in a plane orthogonal to the fixed propagation direction. However, knotted polarization curves can be achieved by interfering electric fields with different frequencies and different propagation directions; this creates the three oscillation directions needed for three-dimensional curves~\cite{sugic2020knotted}. In contrast to direct interference, non-paraxiality can be achieved by tight focusing, where the desired differing propagation directions---and, consequently, all three components of the electric field---naturally arise. In this scenario, a large longitudinal component of the focused field is generated from the radial component of the input field. Meanwhile, the azimuthal contribution remains practically unchanged. As shown in Figure~\ref{fig:fig2}-(a), this is a direct consequence of the action of the lens in transforming the triad $\{\mathbf{e}_r,\mathbf{e}_{\phi},\mathbf{k}\}$~\cite{richards1959electromagnetic}, where $\mathbf{e}_r$, $\mathbf{e}_{\phi}$ and $\mathbf{k}$ represent radial, azimuthal and wavevector unit vectors, respectively.
\begin{figure*}[t]
\centering
{\includegraphics[width=14 cm]{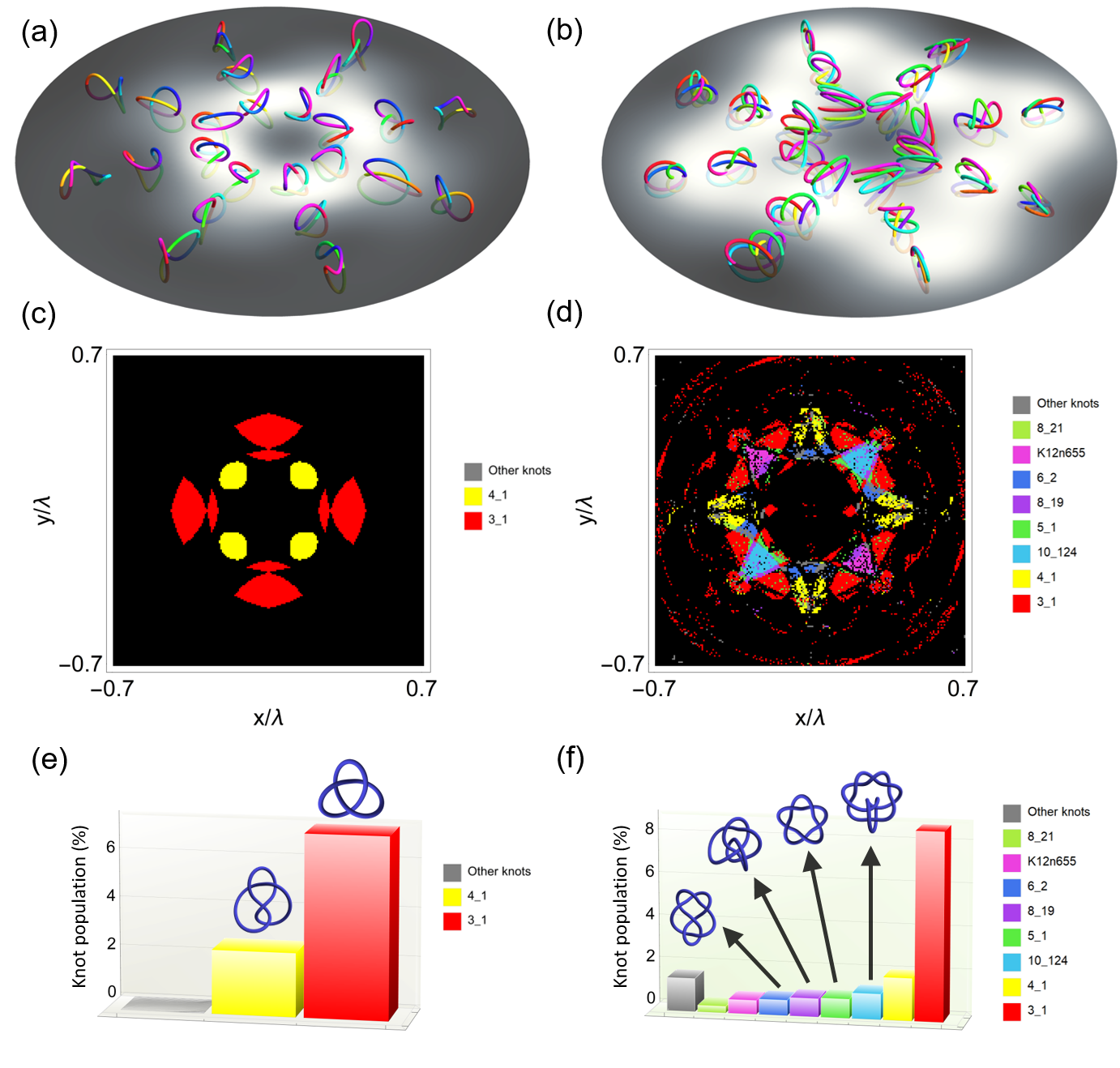}}
\caption{Distribution of polarization curves (forming knots) at the focal plane obtained by tightly focusing the field given by \eqref{Eq:KnotField}. For (a), (c) and (e), $\omega_2/\omega_1=3$ and $\omega_3/\omega_1=2$, and for (b), (d) and (f), $\omega_2/\omega_1=3/2$ and $\omega_3/\omega_1=5/2$. In (a) and (b), some of the curves traced by the local electric field superimposed with the total intensity are shown. For (c) and (d), we performed a point-by-point classification of the knots traced by the local electric field. Different colors correspond to different types of knots, as specified in the legend according to the standard nomenclature. Black color is associated with unknots. The histograms (e-f) show the relative probability density of the different knot topologies within the considered region.}
\label{fig:fig3}
\end{figure*}
In a straightforward manner, it is possible to achieve knots via a  superposition of Gaussian beams (see Supplementary Material); however, this does not produce a large area containing these structures. Instead, specifically structured beams can be chosen such that the three electric field components have similar magnitudes at the focal plane. A possible approach is by looking at the tightly focused polarization patterns of vector vortex beams.

One class of vector vortex beams, in cylindrical coordinates $\mathbf{r}=(r,\phi,z)$ (at the lens input pupil plane), can be defined as,
\begin{equation}
    \mathbf{E_v}(\mathbf{r}\,;\ell,\delta)=\frac{1}{\sqrt{2}} A_{|\ell|}(r) \left[ e^{i\ell\phi}\, \mathbf{\hat{e}_R}+ e^{i \delta}  e^{-i\ell\phi}\, \mathbf{\hat{e}_L} \right].
    \label{eq:VVB}
\end{equation}
Here, $\{\mathbf{\hat{e}_L},\,\mathbf{\hat{e}_R}\}$ are unit vectors for left- and right-handed circular polarization; the factors $\exp{(\pm i\ell\phi)}$---with $\ell$ being an integer---give the OAM content of each circular polarization component; and $A_{| \ell |}(r)$ specifies the complex radial amplitude distribution of the electric field. In our calculations, we considered the case of Laguerre-Gaussian modes,  $A_{| \ell |}(\mathbf{r})e^{i\ell\phi}=\text{LG}_\ell^0(r, \phi)$, with radial index $p=0$ and topological charge $\ell$. The three-dimensional polarization distributions at the focal plane $z=0$ are shown in Fig.~\ref{fig:fig2}-(b-e) for the cases of $|\ell|=1$ and $\delta=\{0,\pi \}$ in \eqref{eq:VVB}. The focused fields are obtained by means of numerical calculation of the Richards-Wolf diffraction integral~\cite{richards1959electromagnetic}. Note that, by considering a polychromatic superposition, i.e.,
\begin{eqnarray}
    \label{Eq:KnotField}
    \mathbf{E}(\mathbf{r},t)&=&a_1 \,e^{-i \omega_1 t} \,\mathbf{E_v}(\mathbf{r},1,0)+a_2\, e^{-i \omega_2t}\,\mathbf{E_v}(\mathbf{r},-1,0)\cr 
    &&+a_3 e^{-i \omega_3t}\,\mathbf{E_v}(\mathbf{r},-1,\pi),
\end{eqnarray}
and focusing it, the curve traced by the local electric field vector takes the form of a Fourier-(3,3,3) knot in its more general form. Additionally, in order to maximize the region in which the longitudinal component of the field is non-negligible, we set $\omega_1=\text{min}(\{\omega_n\})$. This condition arises from the fact that, at higher frequencies, the region where the beam amplitude is non-neglible becomes smaller (see Supplementary Material). Therefore, knotted electric field curves are formed at the focal plane depending on the chosen temporal frequencies and the local relative phase and amplitudes of the chromatic components of the input beam.

\begin{figure*}[t]
\centering
{\includegraphics[width=16 cm]{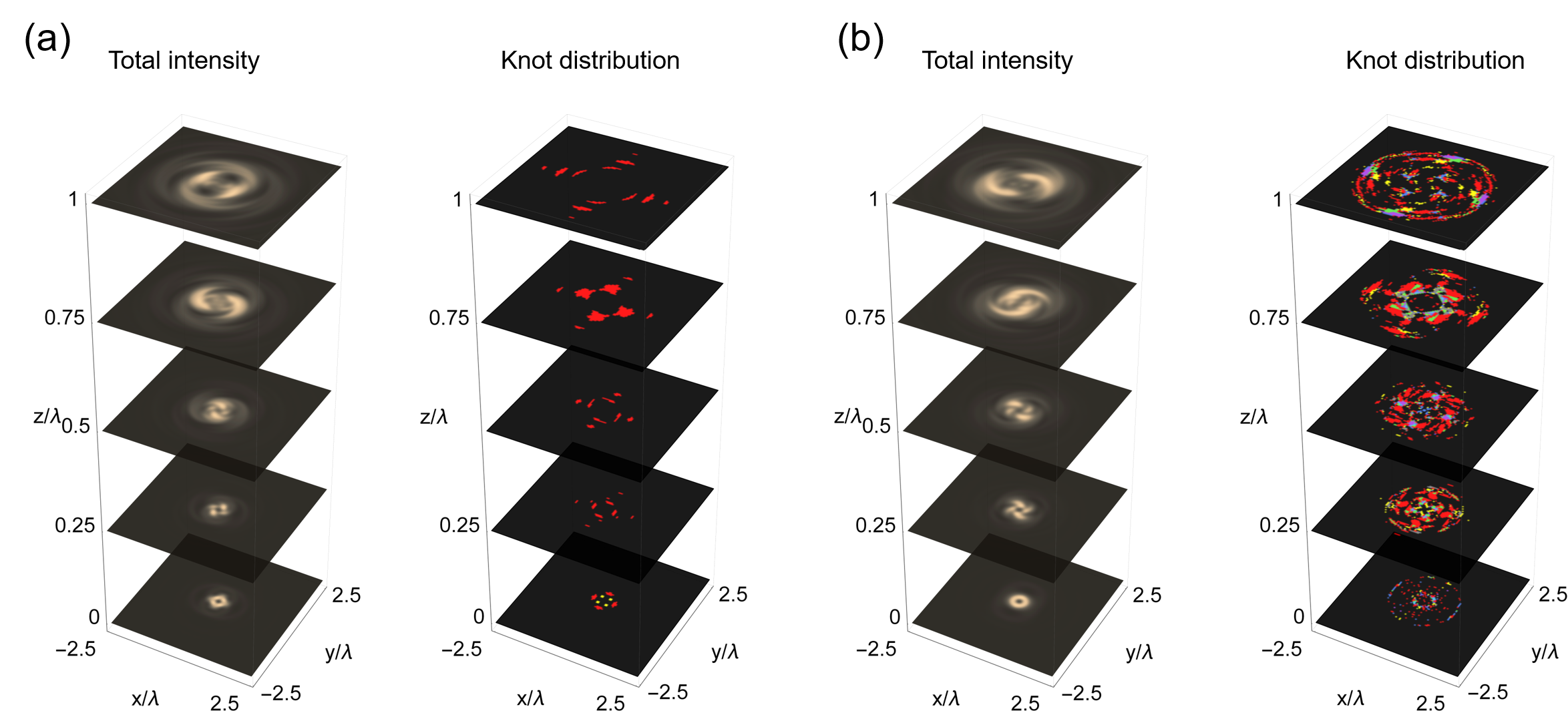}}
\caption{Effect of propagation on total intensity and knotted polarization distribution  for (a) $\omega_2/\omega_1=3$ and $\omega_3/\omega_1=2$, and for (b) $\omega_2/\omega_1=3/2$ and $\omega_3/\omega_1=5/2$. The colour coding in the knot distribution plots is the same as given in Fig. \ref{fig:fig3}.}
\label{fig:fig4}
\end{figure*}

As a first example, let us consider the temporal frequencies in \eqref{Eq:KnotField} to be related by $\omega_2/\omega_1=3$ and $\omega_3/\omega_1=2$, which correspond to the first three co-prime numbers. In addition, we set $a_2/a_1=a_3/a_1=2$ and consider the same beam waist parameter for the Laguerre-Gaussian modes. Figure~\ref{fig:fig3}-(a) shows the three-dimensional polarization distribution as well as the knot distribution and probability density at the focal plane for this particular case. Note that the non-uniform amplitude distribution due to the tight focusing generates a plethora of polychromatic polarization states. While some knots can be recognised by visual inspection, numerical tools were employed for a more systematic approach in order to classify the three-dimensional curves~\cite{pyknotid}. A set of ten thousand curves uniformly distributed within the numerical window at the plane of interest in the focal volume was considered in the classification. Figure~\ref{fig:fig3}-(c) shows the regions in which different knotted structures are located on the focal plane. About $7.5\%$ of the analyzed trajectories were identified as trefoil knots $(3\_1)$, while $2.5\%$ corresponded to figure-eight knots $(4\_1)$ (see Figure~\ref{fig:fig3}-(e)). It should be noted that, by following a circular path around the origin, it is possible to observe the transition between unknots and knotted trajectories (see the Supplementary Movies). Similarly, we present the analysis of a case in which the superposition is composed of structured beams with higher harmonics of an arbitrary frequency. For instance, we consider the frequencies $\omega_2/\omega_1=3/2$ and $\omega_3/\omega_1=5/2$, where these values correspond to the next three lowest possible co-prime numbers: 2, 3 and 5. As shown in Fig.~\ref{fig:fig3}-(b), (d) and (f), about $25\%$ of the analyzed trajectories are non-trivial knots, including the presence of trefoil, cinquefoil $(5\_1)$ and three-twist knots. 

Heretofore, we have considered the existence of knotted polarization states exclusively on the focal plane $(z=0)$. A more complete analysis includes the evolution of the three-dimensional polarization distribution along propagation in the focal volume. As shown in Figure~\ref{fig:fig4}, knotted structures adiabatically follow the beam propagation. In Figure~\ref{fig:fig4}-(a), trefoil knots are present in a region that extends beyond the fundamental wavelength $\lambda=2\pi c/\omega_{min}$. Our calculations show that, since the longitudinal component of the electric field becomes negligible far from the focus, these polarization knots disappear at a distance $\Delta z=2\lambda$ from the focal plane (not shown in the figure). Similarly, Figure~\ref{fig:fig4}-(b) shows the knot propagation for the case $\omega_2/\omega_1=3/2$ and $\omega_3/\omega_1=5/2$.

In summary, we have introduced a heuristic criterion to generate large and rich populations of locally knotted polarization curves by tightly focusing three structured beams at different frequencies. These structures can be generated and detected with the current techniques developed in nanophotonics, e.g., by properly analyzing the scattering of the focused beams from metal nanoparticles from polychromatic beams~\cite{bauer2014nanointerferometric}. In addition, interaction of these locally knotted fields with (semiconducting) materials might induce local three-dimensional magnetic behaviour. 

\vspace{0.2cm}
\noindent \textbf{Funding.} This work was supported by Canada Research Chairs (CRC), Canada First Excellence Research Fund (CFREF), and Ontario's Early Researcher Award.

\vspace{0.2cm}
\noindent \textbf{Acknowledgments.} The authors acknowledge the fruitful conversations with Dr. Sergey Nechayev.

\vspace{0.2cm}
\noindent \textbf{Disclosures} The authors declare no conflicts of interest.

\vspace{0.2cm}
\noindent \textbf{Data availability.} Data underlying the results presented in this paper are not publicly available at this time but may be obtained from the authors upon reasonable request.

\vspace{0.2cm}
\noindent \textbf{Supplemental document.} See Supplement 1 for supporting content.

\bibliography{apssamp}

\begin{figure}[htbp]
\centering
\fbox{\includegraphics[width=15 cm]{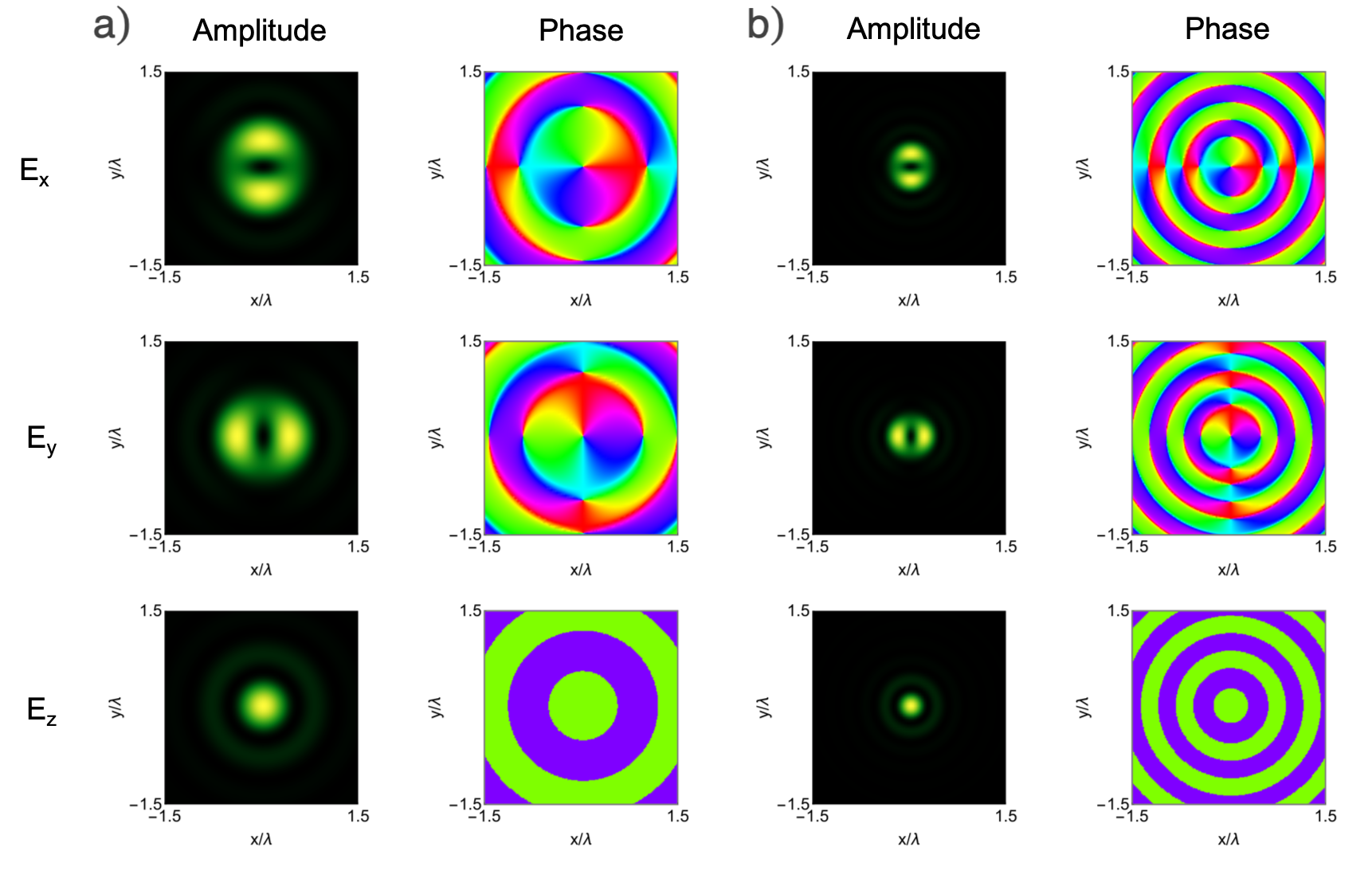}}
\caption{Amplitude and phase distributions of the electric field components obtained by tightly focusing a LG$_{1}^0$ mode (with left circular polarisation). Panels (a) and (b) correspond to two different optical frequencies: (a) $\omega_1=2\pi c/\lambda$; (b) $\omega_2=2 \omega_1$.}
\label{fig:S1}
\end{figure}

\onecolumngrid
\newpage
\vspace{1cm}
\begin{center}
    \Large \textbf{Supplemental Document}
\end{center}
\vspace{0.2cm}
\textbf{Tightly focused polychromatic light.} The time dependent electric field $\mathbf{E}_f(\rho,\varphi,z,t; \omega)$ at the focal plane of an aplanatic lens, with focal length $f$ and a high numerical aperture (NA), is given by Richards and Wolf's diffraction integral~\cite{richard:59},

\begin{equation}
    \label{eq:Efocusing}
    \mathbf{E}_f(\rho,\varphi,z,t; \omega)=-
    \frac{i f\omega}{2\pi c}\,e^{i \omega t}\, \int_{0}^{\Theta}\sin\theta\,
    \mathrm{d}\theta e^{ikz\cos\theta}\int_{0}^{2 \pi} 
    \mathbf{A}(\phi,\theta,t; \omega)
    e^{ik\rho\sin\theta\cos(\varphi-\phi)}\mathrm{d}\phi,
\end{equation}
where $k=\omega/c$ stands for the wavenumber, $\Theta=\text{arcsin}(\text{NA}/n)$ is the maximum angular aperture of the objective, $n$ is the refractive index of the medium, and $\mathbf{A}(\phi,\theta)= \mathbf{T}\cdot\mathbf{E_{\mathrm{in}}}(\phi,\theta) $ is the transformation of the initial beam $\mathbf{E_{\mathrm{in}}}=(U_x,U_y,0)^T$ after the objective. The transformation matrix is given as, 
\begin{equation}
    \mathbf{T}= \sqrt{\cos\theta}\begin{bmatrix} \cos \theta \cos ^2\phi+\sin ^2\phi & (\cos \theta -1)\sin \phi \cos \phi & \sin \theta \cos \phi \\
 (\cos \theta -1)\sin \phi \cos \phi & \cos \theta
   \sin ^2\phi+\cos ^2\phi & \sin \theta \sin \phi \\
 -\sin \theta \cos \phi & -\sin \theta \sin \phi & \cos \theta \end{bmatrix}. 
\end{equation}
Similarly, it is possible rewrite \eqref{eq:Efocusing} to be,  \begin{align}
    \label{eq:EfocusingBessel}
    \mathbf{E}_f(\rho,\varphi,z)=-\frac{if \omega}{2 \pi c} e^{i\omega t}\, \int_{0}^{\theta}\sin\theta\,
    \mathrm{d}\theta e^{ikz\cos\theta}\int_{0}^{2 \pi} 
    \mathbf{A}(\phi,\theta)
   \sum_{m=0}^{\infty} i^{m} \mathrm{J}_{m}(k \rho \sin \theta) e^{im(\varphi-\phi)} \, \mathrm{d}\phi, 
\end{align}
where $\mathrm{J}_{m}(\cdot)$ stands for the $m$-th order Bessel function of the first kind. Therefore, it is possible to construct a tighly focused polychromatic field as, 
\begin{equation}
    \label{eq:Focusing2}
    \mathbf{E}(\rho,\varphi,z,t)=\sum_{m=1}^N \text{Re}\left[a_m\mathbf{E}_f^{m}(\rho,\varphi,z,t; \omega_m)\right].
\end{equation}
where $\{\omega_m\}$ are complex numbers that, 
when looking at the formation of polarization knots, we expect to find these structures in regions where the intensity of the three frequency components are comparable. It is thus useful to keep in mind that the spatial extension of a tight focused beam decreases with the frequency, as illustrated in Fig. \ref{fig:S1}. 

\begin{figure}[htbp]
\centering
\fbox{\includegraphics[width=15 cm]{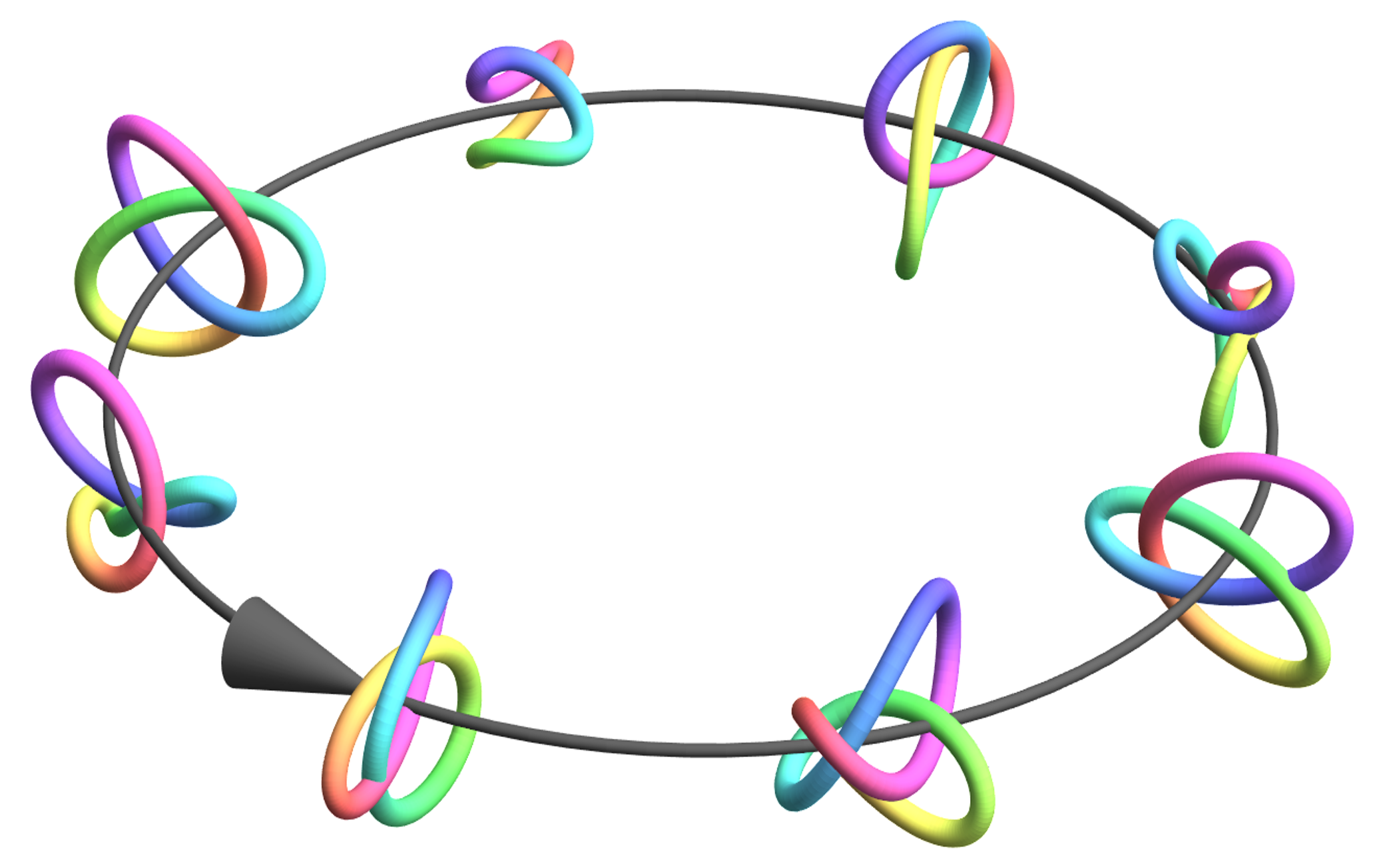}}
\caption{ Evolution of the three-dimensional polarization curves as we follow a trajectory in the focal plane. As we move along path, smooth deformations to the curve occur, leading to transitions points where the curve self intersects and divides the regions where unknots and trefoil knots can be found.}
\label{fig:S2}
\end{figure}

\vspace{0.3cm}
\noindent\textbf{Numerical identification of the knots.} In order to construct the local polarization curves $\mathcal{C}(t)$, each chromatic component of the field is calculated by numerically integrating \eqref{eq:Focusing2}. The identification was performed using \texttt{pyknotid}, a Python module for calculations involving knots or links. In particular, the classification of a curve is based on obtaining the knot determinant~\cite{kauffman2001knots} for each curve and compare it with the built-in database of the module. It should be noted that many of the polarization curves obtained were nearly self-intersecting. This may increase the difficulty of the topology identification, since we did not consider singular knots, i.e. knotted curves that self intersect in one or more points. These self-intersecting curves are a product of the smooth deformation that occurs by following a trajectory between two points on space with different topologies (See Fig.~\ref{fig:S2}).  

\begin{figure}[htbp]
\centering
\fbox{\includegraphics[width= 15 cm]{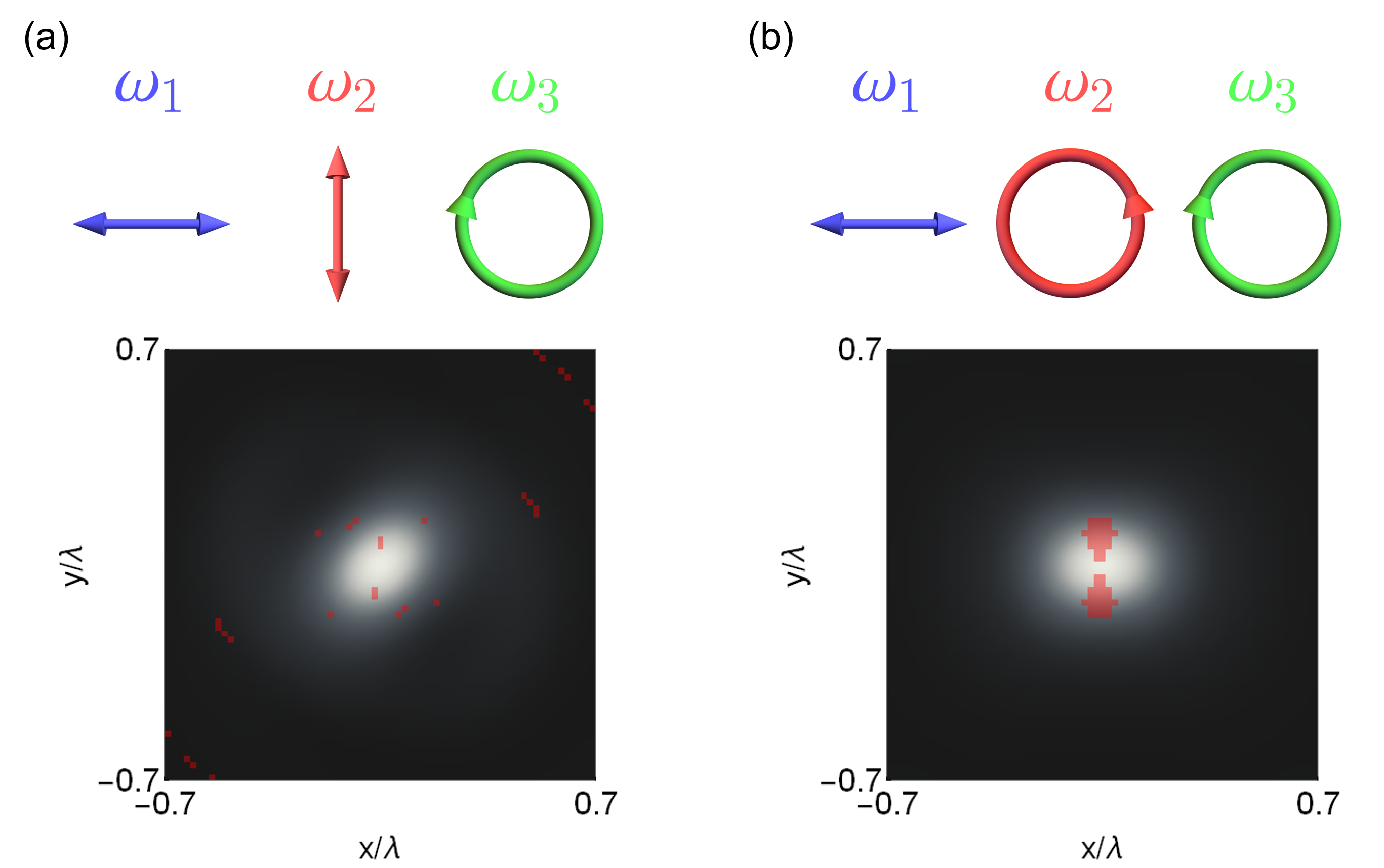}}
\caption{Tight focusing of polychromatic Gaussian beams where each frequency component is prepared in a different polarization state. The frequencies values are given by $ \omega_2/ \omega_1=2$ and $\omega_3/\omega_1=3$ while the relative amplitudes are $a_2/a_1=a_3/a_1=2$. The corresponding polarization state for each frequency are shown above the panels.}
\label{fig:S4}
\end{figure}

\vspace{0.3cm}
\noindent\textbf{Knots formed from tightly focused polychromatic Gaussian beams.} As mentioned in the main Article, knotted polarizations can arise also in the simple case of three Gaussian modes, although the variety of knot types is less rich in this case. In Fig. \ref{fig:S4} we show two examples where each Gaussian mode has a different polarization. Trefoil knots appear in both cases but are more diffused when two modes are circularly polarized (panel b).

\vspace{0.3cm}
\noindent\textbf{Other knot patterns obtained from vector vortex beam superpositions.} Let us consider a family of vector vortex beams, in cylindrical coordinates $\mathbf{r}=(r,\phi,z)$ (at the lens input pupil plane), defined as,
\begin{equation}
    \label{eq:VVB2}
    \mathbf{E_v}(\mathbf{r}\,;\ell,\delta)=\frac{1}{\sqrt{2}} A_{|\ell|}(r) \left[ e^{i\ell\phi}\, \mathbf{\hat{e}_R}+ e^{i \delta}  e^{-i\ell\phi}\, \mathbf{\hat{e}_L} \right].
\end{equation}
Here, $\{\mathbf{\hat{e}_L},\,\mathbf{\hat{e}_R}\}$ are unit vectors for left- and right-handed circular polarization, the factors $\exp{(\pm i\ell\phi)}$, with $\ell$ being an integer, give the OAM content of each circular polarization component; and $A_{|\ell|}(r)$ specifies the complex radial amplitude distribution of the electric field. In our calculations, we considered the case of Laguerre-Gaussian modes, $A_{|\ell|}(\mathbf{r})e^{i\ell\phi}=\text{LG}_\ell^0(r, \phi)$, with radial index $p=0$ and topological charge $\ell$. Let us propose the input field at pupils plane as,
\begin{eqnarray}
    \label{Eq:KnotField2}
    \mathbf{E}(\mathbf{r},t)=a_1 \,e^{-i \omega_1 t} \,\mathbf{E_v}(\mathbf{r};1,0)+a_2\, e^{-i \omega_2t}\,\mathbf{E_v}(\mathbf{r};-1,0)+a_3 e^{-i \omega_3t}\,\mathbf{E_v}(\mathbf{r};-1,\pi).
    \label{eq:VVBsuperposition}
\end{eqnarray}

\begin{figure}[htbp]
\centering
\fbox{\includegraphics[width=15 cm]{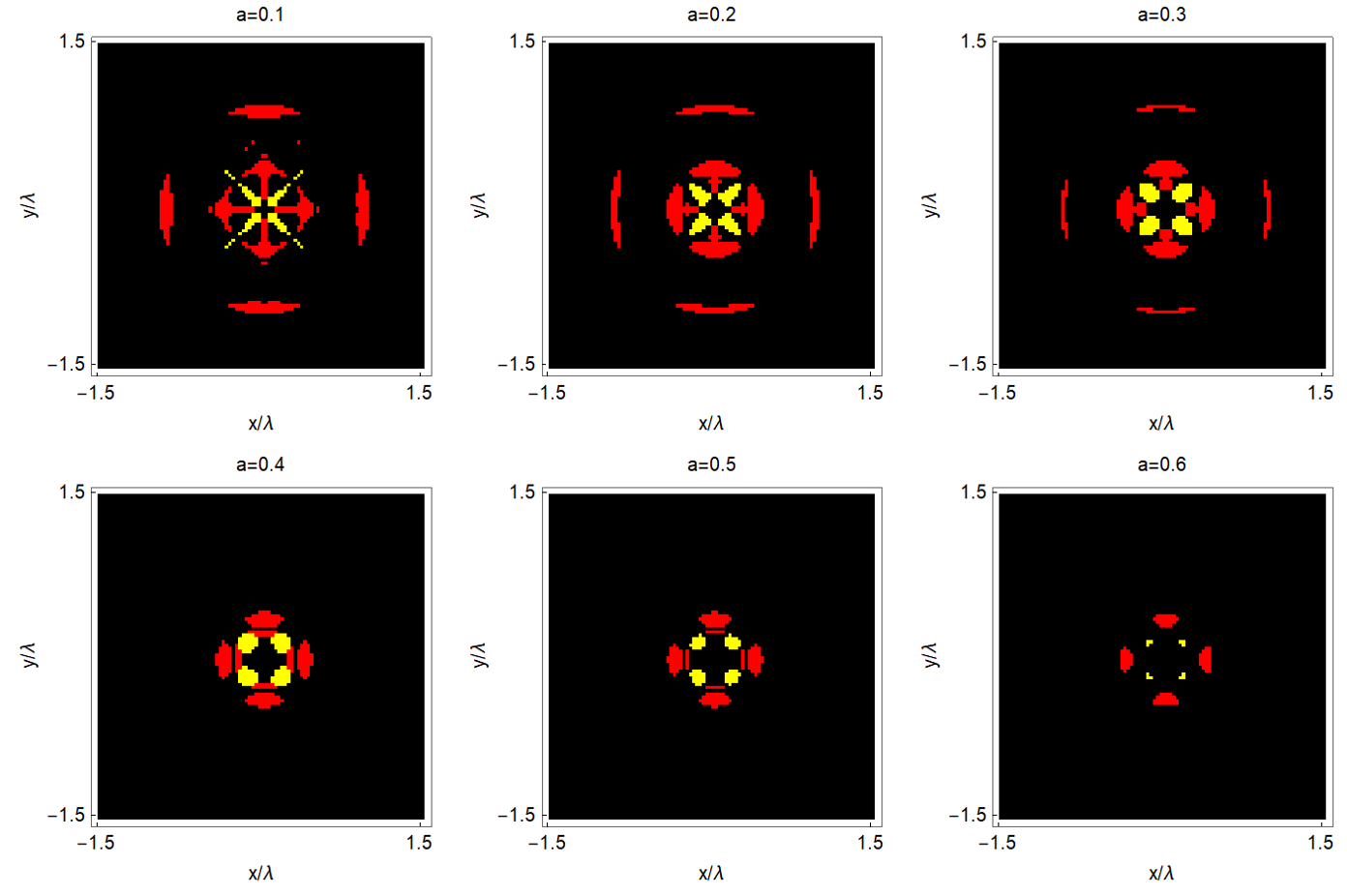}}
\caption{Spatial distribution of knotted polarization curves for when for different values of $a_2/a_1=a$ when the frequencies follow the ratios $\omega_2/\omega_1=3$ $\omega_3/\omega_1=2$.} 
\label{fig:S3}
\end{figure}

\begin{figure}[htbp]
\centering
\fbox{\includegraphics[width=15 cm]{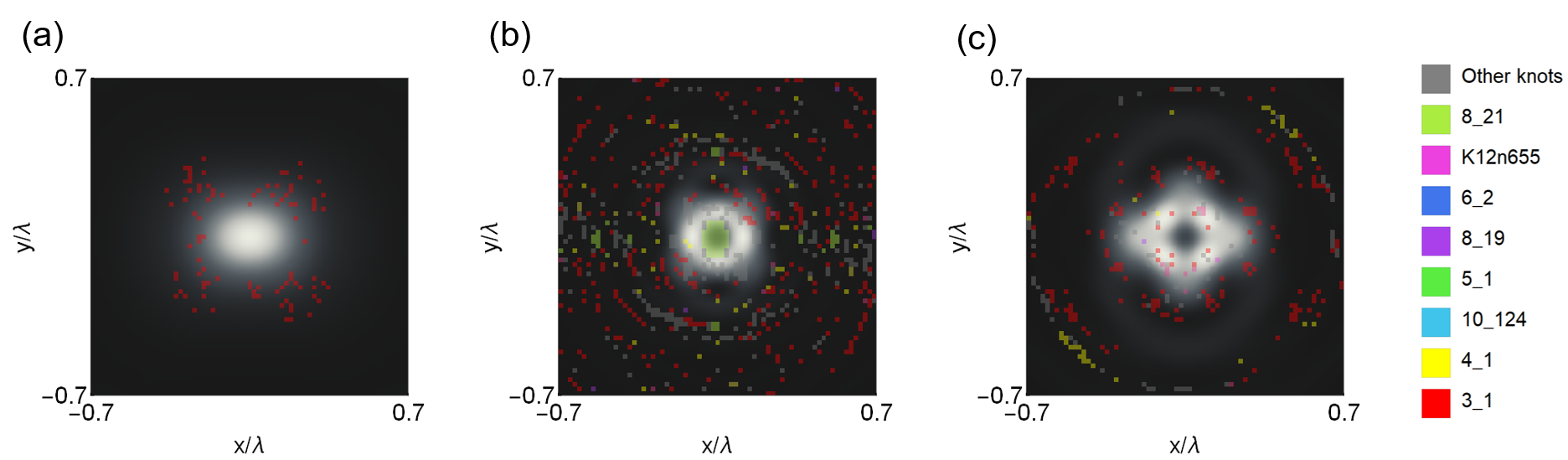}}
\caption{Spatial distribution of knotted polarization curves at the focal plane obtained by tightly focusing the field given by \eqref{Eq:KnotField2}, superimposed with the total intensity for (a) $\omega_2/\omega_1=2$ and $\omega_3/\omega_1=3$,  (b) $\omega_2/\omega_1=5$ and $\omega_3/\omega_1=3$, and (c) $\omega_2/\omega_1=4/3$ and $\omega_3/\omega_1=5/3$. }
\label{fig:S5}
\end{figure}

It is worthy to emphasize that only the non-negligible longitudinal component arises from focusing $\mathbf{E_v}(\mathbf{r},1,0)$, while the remaining terms generate the transverse basis on the focal volume. Therefore, we present a brief exploration on the effect of the relative amplitude of the longitudinal component. For simplicity, we set $a_1/a_2=a_1/a_3=a$ with $a=[0,0.6]$. As shown in Fig. \ref{fig:S3}, in all the cases, both trefoils and figure eight polarization knots are created. It becomes evident that it is possible to use the parameter to tailor the spatial distribution of the knotted polarization curves.
Furthermore, we considered a few additional examples from Eq.~\ref{eq:VVBsuperposition} where we changed only the values of the frequencies $\{\omega_i\}$, keeping the relative amplitudes fixed to $a_2/a_1=a_3/a_1=2$. As shown in Fig.~\ref{fig:S5}, different topologies occur depending on the chosen frequencies and how they are ordered. 

\end{document}